\documentclass[cite1,clbiba,12pt,psfig]{article}
\usepackage{epsfig}
\usepackage{graphicx}
\usepackage{overcite}
\usepackage{chapterbib}

\usepackage{fancyhdr}
\begin{document}

\begin{center}
  
  {Prediction of experimental properties of CO$_2$, improving actual force fields .}
  \vspace{1cm}

  {Ra\'ul Fuentes-Azcatl\footnote{razcatl@xanum.uam.mx} and Hector
    Dom\'inguez$^*$\footnote{hectordc@unam.mx}\\
Instituto de Investigaciones en Materiales,
Universidad Nacional Aut\'onoma de M\'exico,
M\'exico, D.F. 04510 \\ }

\end{center}

\newpage

Abstract

Most of the existing classical CO$_2$ models fail to reproduce some or
many  experimental properties such as surface tension, vapor pressure,
density and dielectric constant at difference thermodynamic
conditions. Therfore, it is proposed a new computational model to
capture better structural, dynamical and thermodynamic properties for
CO$_2$. By scaling the Lennard Jones parameters and point charges;  three
target properties, static dielectric constant, surface tension and
density, were used to fit actual experimental data. Moreover, by
constructing a flexible model, effects of polarization might be
included by variations of the dipole moment. Several tests were
carried out in terms of the vapour-liquid equilibria, surface tensions
and saturated pressures showing  good agreement with experiments.
Dynamical properties were also studied, such as diffusion coefficients
and viscosities at different pressures, and good trends were obtained
with experimental data.

\section{Introduction}

CO$_2$ is the major waste contribution of the  energy production and the
most important greenhouse
gas leading to global climate change. Since the Industrial Revolution
anthropogenic emissions, primarily from fossil fuels and
deforestation, have rapidly increased their concentration in the atmosphere,
driving to global warming. Therefore,
capture and retention of CO$_2$ became matter of several studies,
not only for scientist but also for engineers. From the experimental
point of view several techniques have been developed \cite{wang,lee},
however theoretical approaches have been used, such as computer
simulations \cite{li}, as alternatives to study the phenomena. Therefore,
from the computer simulations perspective is important to have
realible models to produce good result compared with experimental data.

 Nowadays, there are several CO$_2$ models, i.e. force fields,
described in the literature. Murthy et al. \cite{murthy} developed
two- and three-site force fields with Lennard Jones (LJ) interactions
that included a point-quadrupole moment located at the center of mass
of the molecule. M\"oller \cite{moller} and Fischer presented a force
field with two interacting LJ sites with a point-quadrupole, with parameters
that fit experimental data of vapour-liquid equilibrium coexistence (VLEC).
Vrabec et al. \cite{vrabec} reparametrized the model to improve
VLEC. The EPM and EPM2 (elementary physical models), developed
by Harris and Yung \cite{harris}, are CO$_2$ force fields widely used
with LJ sites and point charges in every atom. The Lennard Jones
parameters of EPM model
were obtained through the production of the internal energy
and pressure at temperature of T = 239 K whereas the parameters of the
EPM2 were obtained to reproduce the critical
properties. The TraPPE$_{Flex}$ model developed
by  Potoff and Siepmann \cite{potoff1}  was
obtained to fit the VLEC of the CO$_2$ - propane mixture.
Another common CO$_2$ force field, proposed by  Potoff et al. \cite{potoff2},
was obtained  by modifying the parameters in 
the attractive term of the interaction potential
and it included point charges to adjust the VLEC data.

 On the other hand, Zhang and Duan developed a
force field with three LJ sites and point
charges \cite{zhang, merker1} and Merker et al. \cite{merker2} reported a
force field with three LJ sites with a point-quadrupole
represented by three point charges. That model was  optimized for
VLEC data to obtain 0.4$\%$ deviation from the experiments
in the saturated liquid density and 1.8$\%$ in the vapour pressure.
Persson developed a "One-Center" force field
with a quadrupole for the CO$_2$ with parameters adjusted to the VLEC
data \cite{persson}.
ab initio methods have been also used to construct CO$_2$
force fields, for instance Bock et al.  \cite{bock} proposed
an intermolecular potential with 5 interacting sites
and Bukowski et al. \cite{bukowski} proposed an
intermolecular potential from a perturbation theory, however,
Bratschi et al.\cite{bratschi} concluded that the VLEC behavior is
not accurately represented with those models.

Despite the fact that CO$_2$
is quite polarizable, polarizability effects are not
expected in the thermophysical properties of
pure CO$_2$.
However, in CO$_2$ mixtures with polar components, like water,
the properties may not be exact when using non-polarizable
force fields. Vlcek et al. \cite{vlcek} obtained new optimized parameters
for a CO$_2$ - H$_2$O mixture using SPC/E water
and EPM2 force field, however, the compositions of the CO$_2$-rich phase at
348 K were not adequately represented.
Recently, Orozco et al.\cite{orozco} performed MC simulations
in the Gibbs ensemble to study the
VLEC of the  CO$_2$ - H$_2$O mixture. It was found that non-polarizable
force fields, such as the TraPPE model \cite{potoff2}, have limitations
in the prediction of compositions and densities for steam and liquid
phases. 

 It is worthy mentioning that the existing models, using
molecular dynamics, reproduce
some of the CO$_2$ properties but they fail to reproduce others,
i.e. there is not a CO$_2$  model which captures most of the actual
experimental properties. However,
the EPM2 model using Monte Carlo methods captures several properties such
as densities, VLEC  and vapour pressures but it does not
reproduce surface tension or dielectric constants \cite{harris}. Recently
a polarizable model proposed by Jiang et al. \cite{pagatinop}
with gaussian charges using monte carlo methods reproduced
correctly VLEC and vapour pressures and they obtained
good dynamical properties, however, they did not calculate surface tensions
neither dielectric constants.

 In the present paper we propose a new simple
flexible CO$_2$ force field
to reproduce better thermodynamic, structural and dynamical
experimental data. The interest of having a flexible model is
to study changes in the molecular geometry to search thermodynamic
states due to dipole moment variations. A flexible molecule can be
obtained with changes in the O-C-O angle which could
modify the dipole moment and consequently
producing fluctuations in the dielectric constant.

The remaining of the  paper goes as follows. In section 2  
the new model for the CO$_2$/$\varepsilon$ is introduced. In section 3,
the methodology to obtain of the new  parameters is described,
section 4 shows the simulation
details and  the results are analysed in section 5. Conclusions
are presented in section 6.

\section{The CO$_2$/$\varepsilon$ Model}

 Carbon dioxide, CO$_2$, is a linear and symmetric molecule with zero
dipole moment. The model consists of intra and intermolecular potentials.
In fact, in order to have
a flexible model an harmonic potential was included in the
intramolecular interaction, see figure \ref{fig1}.

 \begin{equation}
\label{eqn1}
 U(\theta)\!=\!\frac{k_{\theta}}{2}(\theta-\theta_0)^2 ,
\end{equation}

where  $\theta$ is the angle O-C-O and $\theta_0$ refers
 to the equilibrium value,  $k_{\theta}$ is the spring constant.

 For the intermolecular potential, between two CO$_2$ molecules,
 the Lennard Jones (LJ) and Coulomb interactions are used,

\begin{equation}
\label{ff}
u(r) = 4\epsilon_{\alpha \beta} 
\left[\left(\frac {\sigma_{\alpha \beta}}{r}\right)^{12}-
  \left (\frac{\sigma_{\alpha \beta}}{r}\right)^6\right] +
\frac{1}{4\pi\epsilon_0}\frac{q_{\alpha} q_{\beta}}{r}
\end{equation}

 where $r$ is the distance between sites $\alpha$ and $\beta$,
$q_\alpha$ is the electric charge of site $\alpha$, $\epsilon_0$ is the
permitivity of vacuum,  $\epsilon_{\alpha \beta}$ is the LJ energy scale
and  $\sigma_{\alpha \beta}$ the repulsive diameter for an $\alpha-\beta$ pair.
The cross interactions between unlike atoms are obtained using
the Lorentz-Berthelot mixing rules,

\begin{equation}
\label{lb}
\sigma_{\alpha\beta} = \left(\frac{\sigma_{\alpha\alpha} +
  \sigma_{\beta\beta} }{2}\right);\hspace{1.0cm} \epsilon_{\alpha\beta} =
\left(\epsilon_{\alpha\alpha} \epsilon_{\beta\beta}\right)^{1/2}
\end{equation}

\begin{table}
\caption{Parameters of the CO$_2$ models considered in this work.}
\label{table2}
\begin{tabular}{|l|ccccccccc|}
\hline\hline
model	&	d$_{OC}$	&	k$_\theta$	&$\theta_{OCO}$	&	$\epsilon _{O-O} $	&	$\sigma_{O-O}$
&	$\epsilon _{C-C} $	&	$\sigma _{C-C}$	&	q$_O$ &	q$_C$	\\

&	\AA	&	kJ/mol rad$^2$	&	deg	&	kJ/mol &	\AA	&	kJ/mol	&	\AA	&	e	& e	\\
\hline
CO$_2/\varepsilon$	&	1.170	&	500	&	180	&	0.03547	&	2.363&	0.220578		&	2.62585	&	-0.6204	&	1.2408	\\
EPM2 	&	1.149	&	1236	&	180	&	0.669335	&	3.033	&	0.233865	&	2.757	&	-0.3256	&	0.6512	\\
TraPPE	&	1.160	&	1236	&	180	&	0.656806	&	3.050	&	0.224478	&	2.800	&	-0.3500	&	0.7000	\\

\hline
\end{tabular}
\end{table}

\section{Parametrization Approach}

 Using molecular dynamics (MD) simulations two CO$_2$ models,
TraPPE$_{Flex}$ (flexible)\cite{orozco} and EPM2 \cite{harris},
were initially tested to evaluate some thermodynamic
properties such as the surface tension. Then, with the method suggested by Alejandre et al. \cite{alj95} the surface tension was  determined, however,the computational results did not agree well with the experiments.
Therefore, in order to improve the actual CO$_2$ models series
of molecular dynamics were conducted to find
a set of parameters of a CO$_2$ flexible model
to reproduce better the experiments.
The procedure started using a CO$_2$
TraPPE$_{Flex}$ force field following the method of Salas et al \cite{salas},
i.e. by modification of the $\epsilon$ and
$\sigma$ Lennard Jones parameters to fit the experimental surface tension and the density. In the same procedure, as Fuentes et al.
reported \cite{KBre,NaCle,spce,tip4pe}, the site charges were also scaled until the static dielectric constant was improved.

The purpose of the present work is also to built 
a flexible model, then the harmonic potential constant
is also parametrized to increase flexibility that is linked to
the polarity of the molecule and the dielectric constant by the dipole moment.

\section{Simulation Details}

 Molecular dynamic simulations were performed using 
GROMACS~\cite{gromacs} software and the new CO$_2$ parameters were
estimated with three target properties,
the static dielectric constant, the surface tension and density.
For the surface tension calculations a  CO$_2$
slab was located in a  parallelepiped cell with
dimensions L$_x$ = L$_y$ = 9.269 nm and L$_z$ = 3L$_x$, i.e. the surface area
was large enough to avoid any finite size effects \cite{Orea} and
5324 molecules were used in the simulations. Then, simulations in the NVT
ensemble with periodic boundary conditions applied in all
three directions were conducted. 
The equations of motions were solved using the
leapfrog algorithm with a time step of 2 fs using the Nose-Hoover
thermostat with a parameter of 1.4 and  LINCS algorithm to keep bond
distances. Electrostatic interactions were handled with the
particle mesh Ewald (PME) method \cite{Essmann} with a grid
space of 0.35 nm and a spline of order 4. The truncation potential
distance was 2.6 nm.
Liquid phase simulations were performed in the isotropic NPT ensemble
with a fixed number of molecules, N=500. The LJ
and the real part of electrostatic interactions were truncated at 1 nm,
and the PME method was used for the long-range electrostatic part with
the following settings: a tolerance of 10$^{-5}$
for the real space contribution, a grid spacing of 0.12 nm and a 4-order
interpolation. The energy and pressure corrections ~\cite{gromacs}
implemented in GROMACS were also applied due
to the use of a finite cut-off for LJ interactions.

\section{Results}

 It is already known that the important parameter to modify the
surface tension is the  $\epsilon_{LJ}$ parameter \cite{salas}.
Therefore, simulations at different $\epsilon_{LJ}$ values,
by scaling all $\epsilon_{LJ}$ parameters, were conducted
until the error with the target property was less than 5.6 \% with respect to the
experimental value. The average components of the pressure tensor were obtained
for 30 ns after an equilibration period of 5 ns.
The corresponding surface 
tension, $\gamma$, for planar interfaces was calculated 
from the mechanical definition \cite{alj95},

\begin{equation}
\label{Ec3}
\gamma  = \frac{1}{2} L_{z}[P_{zz} - \frac{1}{2}(P_{xx} + P_{yy})]
\end{equation} 
 where $P_{\alpha\alpha}$ are the diagonal elements of the microscopic
pressure tensor. The factor 1/2 takes into 
account the two symmetrical interfaces in the system.\\

\begin{figure}[]
\begin{center}
\includegraphics[width=4.2in]{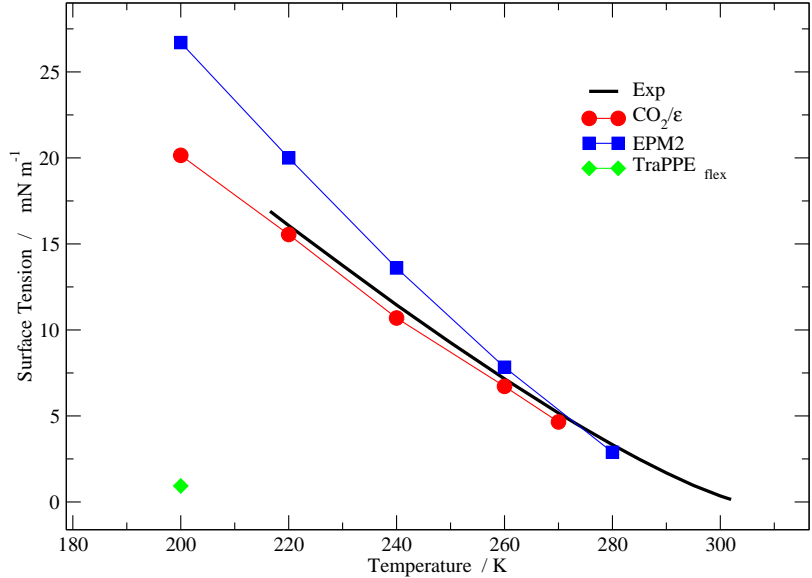}
\caption{Surface tension as a function of temperature for the
  different CO$_{2}$ models. The simulation
  results for the EPM2, TraPPE$_{Flex}$ and
  CO$_{2}$/$\varepsilon$
  were obtained from this work. The continuous line represents the experimental
  data \cite{nist}.}
  \label{fig2}
\end{center}
\end{figure}

In figure \ref{fig2} the surface tension results of the
new CO$_2$/$\varepsilon$
force field with other most common models are shown.
In the figure, it is observed that the new CO$_2$/$\varepsilon$ 
force field describes better the experimental surface tension values 
at different temperatures \cite{nist} . In general, the other CO$_2$
models fail to reproduce the
experiments, only the EPM2 has good agreement with experimental
data between temperatures  T = 260 K and T = 280 K whereas TraPPE$_{Flex}$ 
\cite{orozco} results are even further away from the
experiments.


The selection of the charges in the new model was determined by
calculation of the dielectric constant obtained by the
fluctuations~\cite{neuman} of the total dipole moment
{\bf M},

\begin{equation}
\label{Ec4}
\epsilon=1+\frac{4\pi}{3k_BTV} (<{\bf M}^2>-<{\bf M}>^2)
\end{equation}

 where  $k_B$ is the Boltzmann constant and $T$ is the absolute 
 temperature. The dielectric constant was obtained for long simulations, 40ns,	
 using isotropic NPT ensemble.
The new set of charges were obtained by variation of the original ones,
 of all atomic sites, until the target
 property, the dielectric constant \cite{Moriyoshi}was about \textbf{22.2$\%$ with respect to the
TraPPE$_flex$ and it does not cause other properties to be lost.}

\begin{figure}[]
\begin{center}
\includegraphics[width=3.2in]{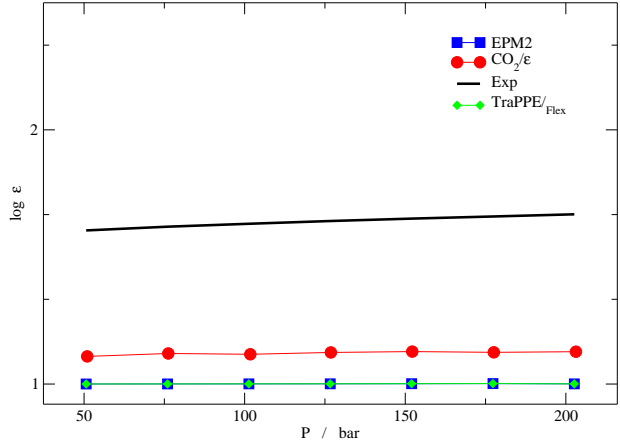}
\caption{Dielectric constant as a function of pressure at 273.15K of
  temperature for
  the CO2/$\varepsilon$, EPM2 and TraPPE force fields. The continuous line
  represents the experimental
data \cite{Moriyoshi}.}
\label{fig3}
\end{center}
\end{figure}

The proper evaluation of the dielectric constant needs long
simulations to have the average dipole moment of the system
around zero. Results of the dielectric constant
at 273.15K of temperature at diferent
pressures are show in the figure \ref{fig3} where
it is observed that the new force field
CO2/$\varepsilon$ shows better agreement with the experiments \cite{Moriyoshi}.

The reproduction of the dielectric constant show that the new force field can modify the CO2 structure in order to capture the change in the dipole
moment.

 With the new $\epsilon_{LJ}$ and charges
values the $\sigma_{LJ}$ of all atoms were also scaled 
to match the  density at 50 bar and 280K of pressure and temperature
repectively with errors less than 17\%.
The new parameters of the CO$_2$/$\varepsilon$ model are
indicated in table \ref{table2}.
Finally, to have a flexible model the angular potential was reduced
from the original one, using an angular constant of k$_{\theta}$ = 500.
This value was chosen since it reproduced better the experimental data.
For all those last calculations simulations in the NPT ensemble
were carried out with different pressures using Nose-Hoover
barostat.

 Flexibility of the model was tested 
at temperature of T = 300 K and pressure of P = 1 bar by measuring the average
O-C-O angle over the simulation time, see table \ref{table3}.
From table \ref{table3} is observed that the new
model is a little more flexible than the others as a consequence of the
reduction of the spring constant, k$_\theta$, in the angular potential.
In fact, it is observed that
the dipole moment increased with respect to the other models
by improving the dielectric constant.
Since fluctuations in the
dipole moment are related 
with the instantaneous polarization then
the factor $G_K$ (equation \ref{Ec5})\cite{Glattli},
was introduced to calculate the differences in the polarization
among the different models studied in the present work,

\begin{equation}
\label{Ec5}
G_K = <{\bf M}^2> / N \mu^2
\end{equation}

 where {\bf M} is the total dipole moment of the system,
N is the number
of molecules and $\mu$ is the dipole moment of a single molecule.
From the results it is depicted that $G_K$ increase for CO$_2$/$\varepsilon$ table \ref{table3}, then the polarizations change in order to improve the dieletric constant figure\ref{fig3}.

\begin{figure}[t]
\centering
\includegraphics[clip,width=6cm]{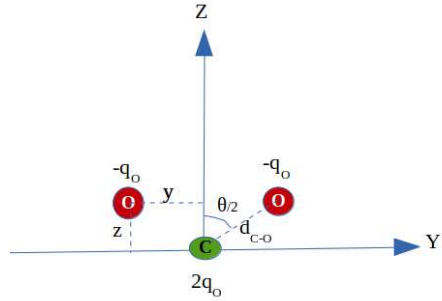}
\caption{Graphical representation of CO2 in gas phase at 300K and 1 bar of temperature and pressure, respectively; According to the data reported in table \ref{table3}}
\label{FigM}
\end{figure}

\textbf{Based on the optimal point charge approximation of Anandakrishnan et al.\cite{QT}
described in figure \ref{FigM} . We calculated the dipole and quadrupole moments, using \ref{Ec1} and \ref{Ec2}, that describe well these dipole and quadrupole moments of the  CO$_2$ molecule in this context. The values are given in table \ref{table3}.}

\begin{equation}
\label{Ec1}
\mu= 2qz 
\end{equation}

\begin{equation}
\label{Ec2}
Q_T= \frac{3qy^2}{2}
\end{equation}

 \textbf{It is observed that the calculated dipole and quadrupole moments of the CO$_2$/$\varepsilon$ model are higher than those obtained with the other models.
Due to the flexibility of CO$_2$/$\varepsilon$ model and the new parameters the electrostatic
moments produce high electrostatic moments.
Even though the CO$_2$/$\varepsilon$ dipolar moment is higher respect to the other
models all values are small. 
On the other hand, the quadrupole moment reported in the
literature is 4.3 D\AA (1D = 0.2082 e\AA) \cite{QTexp} and from table \ref{table3}
is noted that none of the
models have good values, the TraPPE and the EPM2 force fields underestimate
the data with an error of 21 \% and 28 \%, respectively
whereas the CO$_2$/$\varepsilon$  overestimate the value with 42 \% error.
However, in order to have similar quadrupole effects of all
the models, we keep the CO$_2$/$\varepsilon$  quadrupole moment per charge the same as
given in the other models (see last column in table\ref{table3}).
With this approximation not only the dielectric constant is improved but also other thermodynamics properties remained.}

\begin{figure}[]
\begin{center}
\includegraphics[width=4.2in]{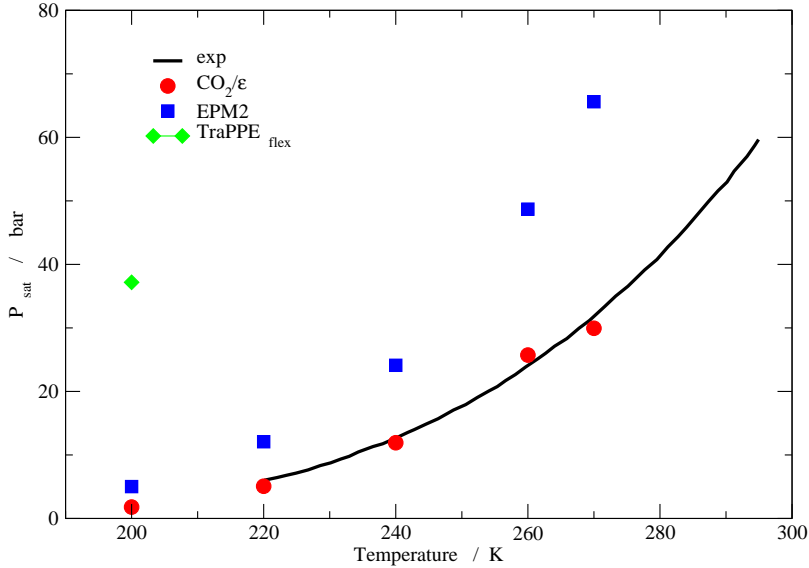}
\caption{Saturated vapour pressures at different temperatures.
  The simulation results
  for the TraPPE$_{Flex}$, EPM2 model and CO$_2$/$\varepsilon$
  obtained from this work. The continuous line represents the experimental
  data \cite{nist}.}
  \label{fig4}
\end{center}
\end{figure}

 One important property within the
chemical engineering community is the calculation of vapour pressures.
In figure \ref{fig4} the results of the vapour pressure
calculated with the perpendicular component of
the pressure tensor respect to the surface are shown
for different CO$_2$ force fields.
It is observed that the new CO$_2$/$\varepsilon$
describes well the vapour pressure , the EPM2 increases its value from
220K and overestimates the experimental value at high temperatures
however the trappe flexible force field overestimate
the experimental values.

\begin{figure}[]
\begin{center}
\includegraphics[width=4.2in]{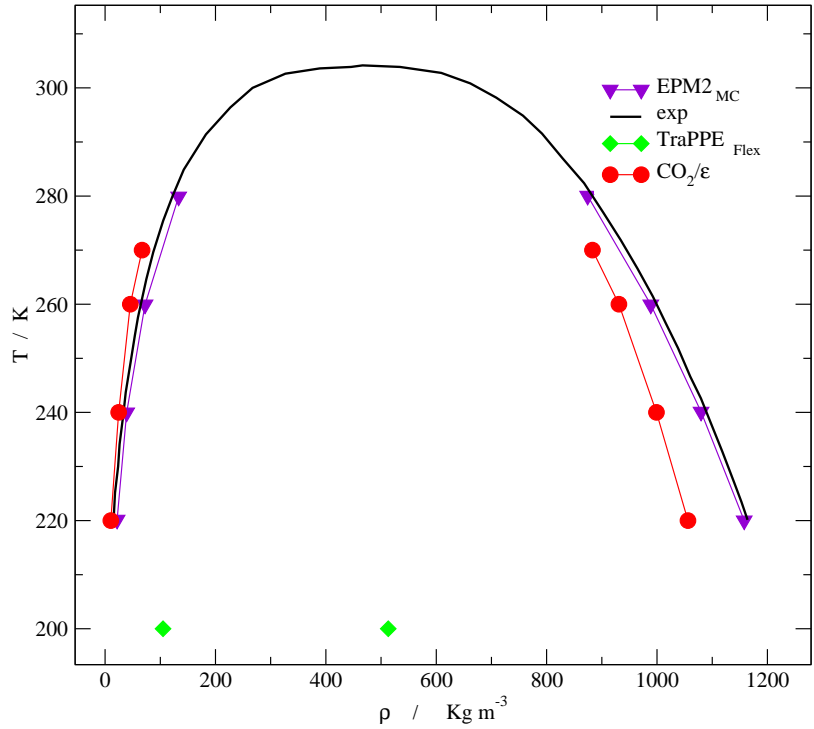}
\caption{Vapour - Liquid phase equilibrium.
  CO$_2$/$\varepsilon$ obtained from this work and the values of EPM2$_{MC}$
  are taken from Harris et al \cite{harris}.   The continuous line represents
  the  experimental data \cite{nist}.}
  \label{fig5}
\end{center}
\end{figure}

 The vapour and liquid densities were calculated with the slab method
described above and the results are plotted in figure \ref{fig5}.
It is noted that the liquid and vapour branches are well described with the
Monte Carlo EPM model \cite{harris}.
The CO$_2 /\varepsilon$ model produces good vapour branch although
the liquid line is not as good as the EPM model.

\begin{table}
  \caption{Results at T = 300 K and P = 1 bar for the CO$_2$ models
    considered in this work.}
\label{table3}
\begin{tabular}{|l|ccccc|}
\hline\hline

	& $\theta_{prom}$	& G$_K$	& Dipole moment	&  $|Q_T|$& $|Q_T/q_O |$ 	\\
	&	$degree$	&		& debye	(D)	& D\AA &	D\AA/e \\
	\hline
TraPPE$_{flex}$		&	179.168	&	1.26086	& 0.0283	& 3.385 & 9.6	\\
EPM2	&	179.174	&	1.24966	&	0.0269 	&3.096& 9.5 \\
CO$_{2}/\varepsilon$		&	178.2&	1.40647 &	0.1095	& 6.117 &	9.8 \\

\hline
\end{tabular}
\end{table}

\newpage

\begin{figure}[]
\begin{center}
  \includegraphics[width=4.5in]{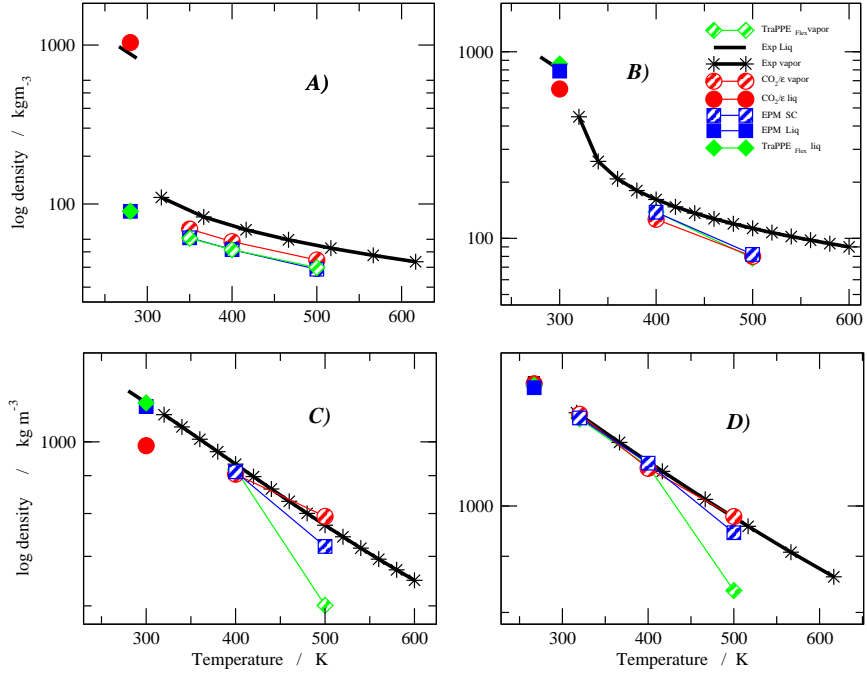}
\caption{Logarithmic of the density as a function of temperature at
  different pressures
A) P = 50 bar.
B) P = 100 bar.
C) P = 1000 bar.
D) P = 2000 bar.
  The simulation results for TraPPE$_{Flex}$ , EPM2 and
  CO$_{2}$/$\varepsilon$ were obtained from this work.
  The solid black line represents the 
  experimental data of the liquid phase and the solid black line
  with stars represents
  experimental data of the vapour phase \cite{nist}.}
  \label{fig6}
\end{center}
\end{figure}

 In order to test the new force field at different
thermodynamic conditions calculations of the density as
function of the temperature at different pressures
are shown in figure \ref{fig6}. 
As a general trend the  CO$_2$/$\varepsilon$ capture the liquid branch
reasonably well at high pressures. However, at high temperatures different
issues are depicted.

\begin{figure}[]
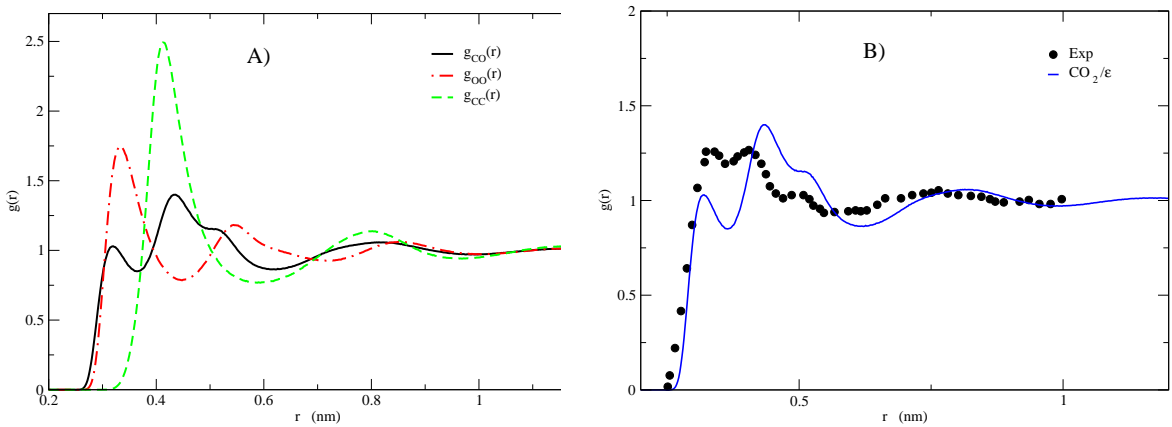

\begin{center}
\includegraphics[width=3.1in]{fig7a.eps}\includegraphics[width=3in]{fig7b.eps}
\caption{Figure A. The  Radial distribution functions g$_{OO}$, g$_{OC}$ and g$_{CC}$ for   CO$_{2}$/$\varepsilon$  are ploted from the calculation of molecular Dynamics at T = 239 K and P = 14.5 bar.
Figure B. Radial distribution functions, g(r), for
  CO$_{2}$/$\varepsilon$ and
  neutron weighted pair correlation function
  at T = 239 K and P = 14.5 bar; experimental pair correlation
function is shown in black circles \cite{Tricht}.
}
\label{fig7}
\end{center}
\end{figure}

 The structure of CO$_{2}$ is represented by the
proposed  CO$_{2}$/$\varepsilon$ force field and plotted in terms of
the pair distribution function (g(r)) in 
figure \ref{fig7}. 
It is observed that the CO$_{2}$/$\varepsilon$ captures better
the first peak indicated by the experiments, i.e.
at 0.319 nm very close to the experimental data, 0.332 nm.
In fact, there is a second peak around 0.434 nm close to the experimental
value of 0.405 nm, figure 8B, \textbf{ is noted the second peak, in the g(r), higher than the first one
whereas the experiments show two peaks nearly at the same height.
Due to the flexibility of the model it is possible that
the CO2 molecule bents more and more attraction between O-O atoms 
compare with the C-O interactions could
be produced by increasing the
second peak of the g(r), i.e the larger number of second nearest neighbors.
At large distances the g(r) of the CO$_{2}$/$\varepsilon$  looks more similar than
the experimental one.}

Experiments of the liquid CO$_{2}$ structure were obtained from
van Tricht et al. using neutron diffraction experiments \cite{Tricht}.

\begin{figure}[]
\begin{center}
\includegraphics[width=4.2in]{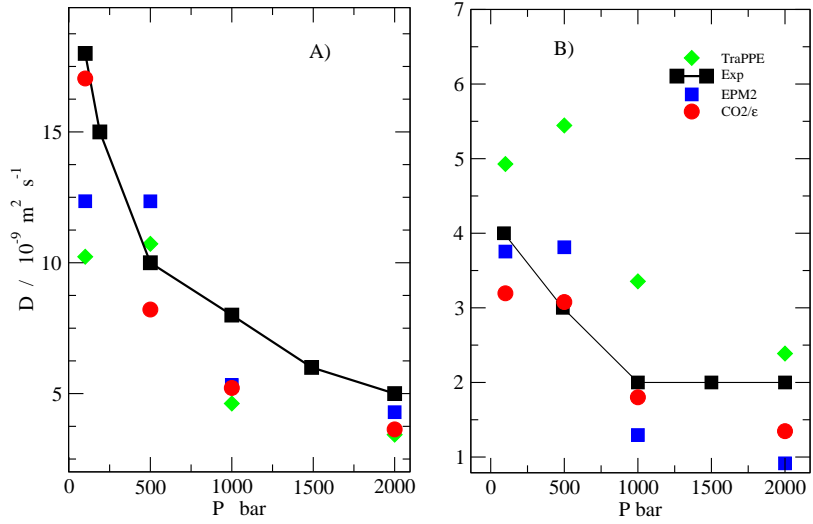}
\caption{Diffusion coefficients  as a function of
  pressure at A)298 K of temperature and B) 223 K of temperature. The simulation results for the TraPPE,
  EPM2 and
  CO$_{2}$/$\varepsilon$ are obtained
  from this work. The black
  square and solid black lines represent the experimental data \cite{nist}.}
  \label{fig8}
\end{center}
\end{figure}

\begin{figure}[]
\begin{center}
\includegraphics[width=4.2in]{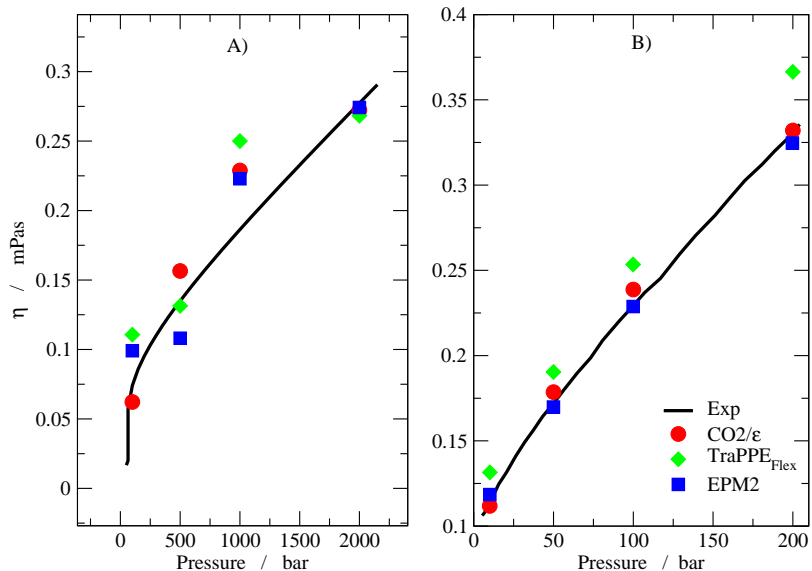}
\caption{Shear viscosity  as a function of pressure at temperature
 A) T = 298K and B) 223K for the TraPPE, EPM2 and CO$_{2}$/$\varepsilon$ models are
  compare. The solid black line represents the experimental data \cite{nist}.}
  \label{fig9}
\end{center}
\end{figure}

 Dynamical properties were also calculated with the new
CO$_{2}$/$\varepsilon$ model.
In figure \ref{fig8},\ref{fig9} results of the diffusion coefficients
as function of the pressure at two different temperatures are shown.
At T = 223 K  the CO$_{2}$/$\varepsilon$
in general reproduces better the values better than
the others models repect to the experiments. At T = 298 K  the 
CO$_{2}$/$\varepsilon$ captures well the shape of the experimental data,
in particular at low pressures, however
data for all models are not good compared with the experiments.

 It is important to mention that the size of the simulated system
for the diffusion coefficients calculations
were big enough to avoid any size effects as indicated in previous
works \cite{Dunweg}.
The diffusion coefficient was obtained from the long-time
limit of the mean square displacement according to the Einstein
relation \cite{Allen},

\begin{equation}
\label{EcD}
D= \lim _{x\to\infty} <(r(t)-r(0))^2> / 6t
\end{equation}

 where r(t) corresponds to the position vector of the center of
mass at time t and the averaging $<...>$ is performed over both
time origins.

 Finally, another dynamical property was also calculated, the
viscosity ($\eta$), and plotted in figure \ref{fig9}. The viscosity
was calculated with the Green–Kubo formula \ref{Ec7} relates the shear
viscosity to
the autocorrelation function ACF of the off-diagonal components of the
stress tensor P$_{\alpha,\beta}$, namely,

\begin{equation}
\label{Ec7}
\eta= \frac{V}{k_BT} \int_0^\infty <P_{\alpha\beta}(t_0)P_{\alpha\beta}(t_0+t)>_{t_0}dt,
\end{equation}

In figure \ref{fig9} is observed that the CO$_{2}$/$\varepsilon$ model
gives the correct tendency with the experiments.

\newpage
\section{Conclusions}

Comprehensive molecular dynamics simulations were conducted to built
a new flexible CO$_{2}$ molecule. Then, series of simulations were carried
out to find a new set of parameters for the CO$_{2}$/$\varepsilon$ model.
The proposed flexible force field captured polarization of the molecule
and reproduced correctly the experimental  
surface tensions, vapor pressures, densities and the static dielectric
constants.
Dynamical properties were also calculated, and even though the CO$_{2}$/$\varepsilon$ model was not parameterized to reproduce these transport properties, the calculations with the new model are close to the experimental values

Variation on the $\epsilon$-LJ parameter influence
interaction between molecules and somehow internal
energy of the system, therefore
thermodynamic properties
might be modified and it could be the reason why the 
parameter is the one that correct the surface tension. On the other hand,
$\sigma$-LJ affects the size estructure and consequently the
system density. Electrostatic properties can be altered
by point charges and therefore properties such as the
static dielectic constant can be modified by adjustment of those
charges. It has been shown that changes of LJ-parameters and charges
can directly affect some properties, however they
also might influence any other by spoiling  correct data
and it could be the reason that in some case some properties
fits correctly whereas others fails in other force fields. \textbf{In fact, by adjusting parameters is hard to reproduce all thermodynamic
and dynamical
properties of the CO2, if some
properties are correctly fitted there are other properties that might be wrong
and vice versa, therefore the best selection of parameters are those
which capture most of the properties with reasonable experimental error.}

Previos works, with using polarizable models \cite{pagatinop} have also shown good results with actual
  experiments, although they do not calculate
  surfacte tensions. Comparisons of the present
  results with a polarizable model are shown in
  the supplementary material. There, it is observed that our results agree
with those data.

On the other hand, force fields using
  classical potentials fail to reproduce some or
many  experimental properties such as surface tension, vapor pressure,
density and dielectric constant at difference thermodynamic
conditions.
Therefore, in the present work, a new CO$_{2}$ force field has been proposed ;
which reproduces
several thermodynamics, structural and dynamical properties 
improving the present classical CO$_{2}$ force fields.

It is worthy to mention that approprite CO2 force fields can help
us to perform more realistic and reliable simulations of actual 
system and to understand better the behaviour of real phenomena such
as CO2 capture where good models are needed to explore real
experiments.

\section {Conflicts of interest}

 There are no conflicts of interest to declare

\section {Acknowledgments}

 The authors acknowledge support from DGAPA-UNAM-Mexico
grant IN102017 and DGTIC-UNAM grant LANCAD-UNAM-DGTIC-238 for
the supercomputer facilities. RFA thanks DGAPA-UNAM for a posdoctoral
fellowship. We also acknowledge Alberto Lopez-Vivas and
Alejandro Pompa for technical support. We also want to thank
the reviewers for their comments of the manuscript, they help to improve
significantly the paper.

\end{document}